\documentclass{article}
\usepackage[utf8]{inputenc}
\usepackage[a4paper, margin=2cm]{geometry}
\usepackage{authblk}  
\usepackage{graphicx}  
\usepackage{caption} 
\usepackage{caption}
\usepackage{wrapfig}
\usepackage{indentfirst}

\author[1]{Q. Cojean-Palassoé}
\author[1]{A. Bertoldi}
\author[2]{A. Landragin}
\author[1]{B. Canuel}

\affil[1]{\textit{LP2N, Laboratoire Photonique, Numérique et Nanosciences,} \\
Université Bordeaux-IOGS-CNRS:UMR 5298, rue F. Mitterrand, \\
F-33400 Talence, France}

\affil[2]{\textit{LTE, Observatoire de Paris, Université PSL, Sorbonne Université, Université de Lille, LNE, CNRS}\\
61 avenue de l’Observatoire, 75014 Paris, France}

\title{Optimizing NN reduction in an atom interferometer network for GW detection}
\date{} 

\begin{document}

\maketitle

The sensitivity of an atom gradiometer aiming to detect gravitational waves (GW) is impacted by fluctuations of Earth's gravity field also called Newtonian Noise (NN) \cite{canuel_elgareuropean_2020}\cite{junca_characterizing_2019}.  Sensor arrays have proved to be a promising technique for NN reduction \cite{chaibi_low_2016}.  In our study, we further investigate the benefits of Atom Interferometer (AI) networks by improving their geometry and the extraction of the GW signal. We focus on Seismic Newtonian Noise in the frequency band from $0.1$ to $10$ Hz. On one hand, we show that using a specific detector geometry, a better NN rejection can occur optimizing the number of gradiometers in the network. On the other hand, we show that carrying out optimization in sub frequency bands -- which results in using various detector geometries from a common network -- allows even higher NN rejection while keeping a similar number of interferometers.

\subsubsection*{Introduction}

\indent It has been shown in \cite{chaibi_low_2016}, that summing the differential phase obtained by gradiometers' measurements enables a NN reduction proportional to $\sqrt{N}$ (N being the number of gradiometers) and even more when benefiting from  the spatial correlations of the NN. In this publication, Chaibi and al. used 80 gradiometers of 16.3 km length, separated by 200 m which equals a total baseline of 32 km. In the following, we will define this configuration as a homogeneous configuration described by a triplet (N,d,L) $\rightarrow $ (80,200,16300). Our study aims to develop numerical tools to improve the performance of homogeneous detector geometries carrying out the optimization in a broad frequency band or dividing the frequency band in sub-frequency bands (see Methodology). To illustrate the gain enhanced by this optimization, the performance reached in Chaibi and al. has been taken as reference and will be referred as ``REF" in the following. For this purpose, we set in our study the maximum number of gradiometers to $N_{max} = 80$ and the total baseline to $L_{tot} = (N-1)\cdot d + L = 32$ km.

\subsubsection*{Methodology}

\begin{wrapfigure}{r}{0.4\textwidth}
    \centering
    \includegraphics[trim=0.20cm 0.10cm 0.2cm 0.05cm, clip,width=0.38\textwidth]{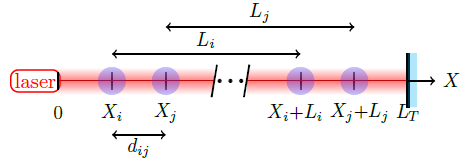}
    \captionsetup{justification=centering}
    \caption{Geometry of the atom gradiometer network.}
    \label{fig:Network illustration}
\end{wrapfigure}In the following, we focus on the NN contribution to the final GW strain sensitivity, and more specifically on Seismic Newtonian Noise (SNN) caused by Rayleigh Surface Waves \cite{harms_low-frequency_2013} which scale as the ratio we optimize in our study: $Le(w) = \chi_N/(\sum_{i=1}^N L_i)^2$, where $Li$ is the respective length of each gradiometer (see Figure~\ref{fig:Network illustration}) and $\chi_N$ gathers all the correlation effects between the various sensors along the network (see Eq.~\ref{eq:chi}). In Eq.~\ref{eq:chi}, $d_{ij}$ corresponds to the distance between $X_i$ and $X_j$ referred in Figure~\ref{fig:Network illustration}, $k$ is the wave number of the seismic wave and $C(d_{ij},k)= \frac{1}{2}[J_0(k\cdot d_{ij})-J_2(k\cdot d_{ij})]$ is the correlation value of the acceleration between two test masses distanced by $d_{ij}$ \cite{junca_characterizing_2019}.

\begin{equation}
\chi_N= \sum_{i,j}
C(d_{ij},k) 
- C(d_{ij} - L_j,k) 
- C(d_{ij} + L_i,k) 
+ C(d_{ij} + L_i - L_j,k)
\label{eq:chi}
\end{equation}

To find the best positioning of sensors, we develop a code which evaluates for every configuration the sum of $Le(w_i)$ ratios and return the geometry minimizing the cost function $\sum_{i=1}^{n}Le(w_{i}) / Le(w_{i})^{REF}$ where $i$ is the discrete frequency number within the investigated band. To minimize the configuration space, the spatial resolution itself has to be discretized by a so called grid resolution $\delta$, i.e. the minimal space between two successive interferometers. Taking the numerical and geometrical constraints, the code returns all possible (N,d,L) given $\delta$, $N_{max}$ and $L_{tot}$, before computing the cost function detailed above. We distinguish two distinct methods namely ``broadband" and ``sub-frequency band" optimization. Broadband optimization consists in integrating $Le(w)$ between 0.1 and 10 Hz which entirely covers the frequency band under study. For sub-frequency band optimization, we divide this frequency band in n sub-frequency bands: $f_1,...,f_n$. The code returns the optimal geometry in the various frequency bands. 

\subsubsection*{Results and Perspectives}

\begin{figure}[h!]
    \centering
    \includegraphics[width=0.5\textwidth]{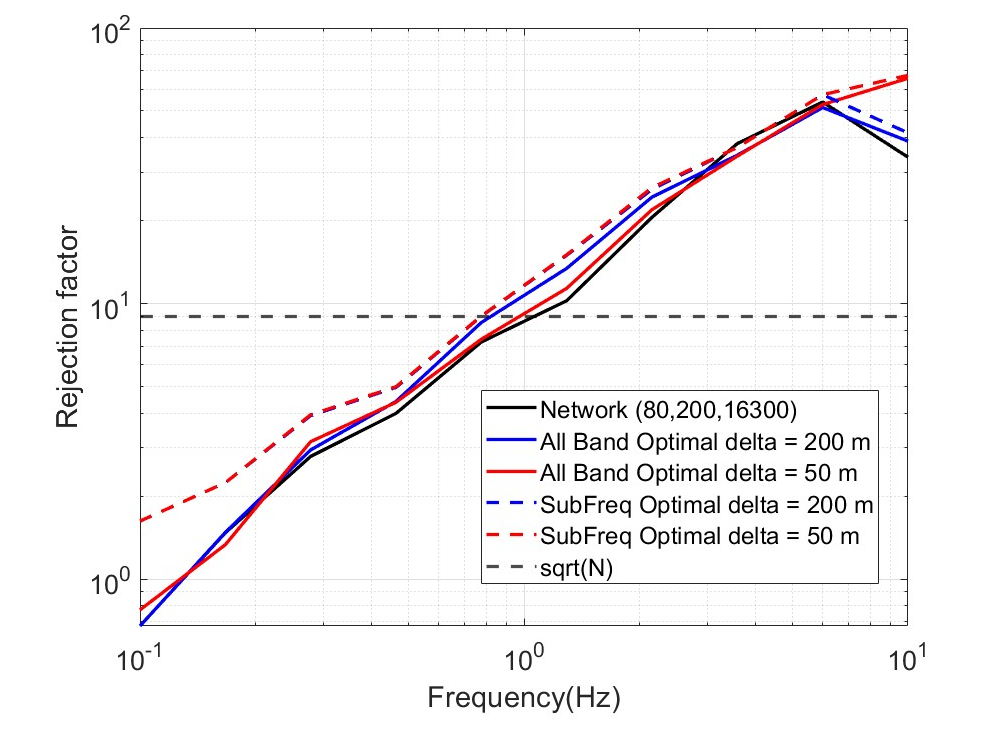}
    \hfill
    \centering
    \includegraphics[width=0.45\textwidth]{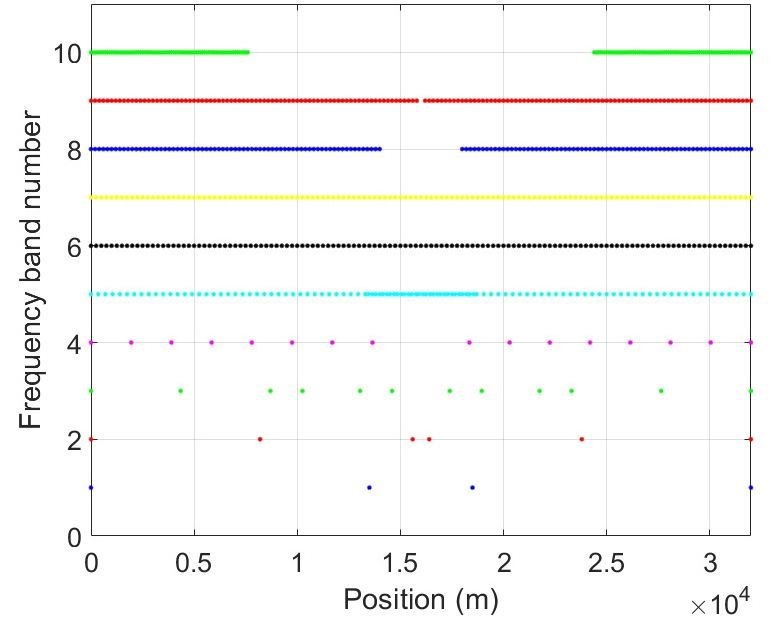}
    \caption{(left) Rejection factor: N gradiometers vs. Single gradiometer $L_{tot}$ for the reference configuration (black line), the optimized network on the full bandwidth with $\delta = 200$ (blue curve) and $\delta = 50$ (red curve) and for the sub-frequency band optimization results (dashed curves). (right) Optimal position of the sensors at each frequency with $\delta = 50 $.}
    \label{fig:homogeneous optimization}
\end{figure}

Fig.~\ref{fig:homogeneous optimization}~(left) compares the SNN rejection factor obtained when doing broadband as well as sub-frequency band geometry optimization for two grid resolutions $\delta = 200$ m and $\delta = 50$ m with the one of the reference configuration. The rejection factor is obtained by the ratio: $\sqrt{Le^1(w)/Le^N(w)}$, $Le^1(w)$ being the NN scale factor for a single gradiometer of length $L=L_{tot}$. For the sub frequency band optimization, Fig.~\ref{fig:homogeneous optimization}~(right) displays the optimal position of sensors along the 32 km for the different frequencies $f_1$ to $f_{10}$ $\rightarrow $  $ \approx (0.1,0.2,0.3,0.5,0.8,1.3,2.2,6.0,10)$ Hz, discretizing the frequency space.

For broadband optimization, with $\delta = 200$ m, we found that the optimal geometry is for (76,200,17000). In comparison with the reference geometry, it has four less gradiometers and an increased single gradiometer baseline $L$. We also studied how grid resolution could improve the performance. We found that going to higher grid resolution, from 200 m to 50 m, can lead to significant improvements above 4 Hz at the cost of a worse rejection factor at 1 Hz. This is due to the cost function which optimizes the rejection in all the detection band at once. This is why we also studied an optimization in sub-frequency bands. 

This last method demonstrates significant improvements in SNN rejection with about half an order of magnitude at lower frequencies. As shown in Fig.~\ref{fig:homogeneous optimization} (right), we obtain different optimal detector geometries in each frequency band. However, we demonstrate that almost all these geometries are coming from a single network. For  $\delta = 200 $ m, we need 161 sensors while the reference geometry is using 160 sensors ($N = 80 $). In comparison with reference \cite{chaibi_low_2016}, we therefore demonstrate a significant improvement of NN rejection by using this optimization in sub frequency bands.

Future work could consist in developing genetic algorithm tools to study higher SNN rejection from inhomogeneous geometries, exploring more theoretical and numerical tools to filter NN and, finally, extend the work in 2D and even 3D. 

\bibliographystyle{iopart-num}
\bibliography{reference}

\end{document}